# Being at Home in the Metaverse? Prospectus for a Social Imaginary

Tim Gorichanaz
Drexel University, Philadelphia, PA, USA
gorichanaz@drexel.edu

**Abstract**: The metaverse has seen growing corporate and popular interest over the past few years. While visions vary, the metaverse is generally seen as an extension of the internet that may be developed through advances in a number of digital technologies, such as augmented and virtual reality, as well as new technical infrastructure and standards. The metaverse constitutes an emerging social imaginary, a way of both understanding and directing our shared existence. This paper examines this emerging social imaginary through the phenomenological concept of dwelling, or being at home in the world, as developed by Martin Heidegger. To examine in depth one influential articulation of this social imaginary, this paper focuses on the metaverse as envisioned by Mark Zuckerberg, CEO of Meta (formerly Facebook). The paper presents a thematic analysis of Zuckerberg's public statements regarding the metaverse to provide a close reading of this particular vision. Then, through the lens of Heidegger's philosophy of dwelling, this paper identifies numerous threats to dwelling posed by the metaverse social imaginary. This paper explains these threats and their prognoses, and it closes with some considerations for how the metaverse could be designed to better facilitate human dwelling.

# Introduction

The word "metaverse" dive-bombed into the vernacular on October 28, 2021. On that day, Mark Zuckerberg announced that his company was changing its name from Facebook to Meta to better reflect its aspirations as "a metaverse company." Zuckerberg first publicly shared these aspirations in a July 2021 interview with *The Verge* and then on Facebook's quarterly earnings call a week later (Chayka, 2021; Matney & Hatmaker, 2021; Newton, 2021). On that call, he described the metaverse as "a virtual environment where you can be present with people in digital spaces. You can kind of think of this as an embodied internet that you're inside of rather than just looking at. … In many ways the metaverse is the ultimate expression of social technology." This paper explores that claim, contextualizing its meaning for our human future.

The discussion in this paper centers on the phenomenological concept of dwelling (Heidegger, 1971), which refers to our sense of feeling at home in the world in which we find ourselves. Dwelling involves our environment as well as our technologies; we humans make and use technologies as an essential part of our being. As the environment and our technologies change, then our dwelling may come under threat and perhaps must adapt. In his time, Heidegger (1971) pointed to threats to human dwelling; today, we experience climate change as well as ongoing advances in technology as possible threats to dwelling. In this paper, I look specifically at the threats to and prospects for dwelling presented by the metaverse. To help us plumb these questions, the metaverse is framed as an emerging social imaginary, or a way of understanding and creating our social and technological circumstances.

This paper proceeds as follows. The next section presents Heidegger's (1971) theory of dwelling, and the following section discusses the concept of the social imaginary. Next, the metaverse is framed as an emerging social imaginary, and then an empirical thematic analysis of the characteristics of this social imaginary is presented. After this, our prospects for dwelling, or a healthy "onlife" existence, with the metaverse are examined. As it is currently being discussed, the metaverse seems to constitute primarily a turning away from proper dwelling rather than a turn toward it. However, as the metaverse social imaginary is still emerging, there are opportunities to shepherd its development toward ethically better ends. The final section of the paper discusses some of these opportunities. Throughout, this paper draws primarily on the work of thinkers in the phenomenological tradition, namely Martin Heidegger, Hannah Arendt and Charles Taylor.

# Dwelling in the 21st Century

In "Building, Dwelling, Thinking," Heidegger (1971) describes the way of being of human beings as *dwelling*. In its everyday sense, the word "dwelling" means taking up residence or spending time in a place; as a noun, the word can also refer to the place where we reside. For Heidegger, dwelling is not just about material survival, but about being at home in the world. According to Heidegger's etymological analysis, the original character of the idea of dwelling had to do with preserving and safeguarding, of not only keeping from harm but of bringing something back "specifically to its being… into a preserve of peace" (p. 147). For Heidegger, this goes in both directions: humans are preserved and guarded by the world just as we (should) preserve and guard it.

For Heidegger, places where we dwell, such as buildings and other sites, are gatherings of four related elements that Heidegger calls "the fourfold." The fourfold come in two pairs: earth and sky, which have to do with space and provision; and mortals and divinities, which have to do with time and goals. The elements can be defined in this way:

- **Earth** is our supporting ground, which cares and provides for us in terms of food, shelter, materials, etc. Perceptually it recedes from us, such that we tend not to notice it, taking its support for granted.
- **Sky** is the expansiveness spread over the earth, complete with its changing seasons and weather and its cycles of night and day. The sky provides a sense of something beyond.
- **Mortals** are the kind of being we are: self-aware about the fact that we die. This awareness is not a cause for despair but rather helps bring into focus the projects we have for our lives.
- **Divinities** are the possibilities pregnant in the things around us mortals; divinities are the meanings we sometimes glimpse in moments of revelation, which offer us vocations and aspirations to strive for.

For Heidegger, we humans dwell insomuch as we save and foster the earth, accept the sky's changing offerings, cultivate hope for the benevolence of the divinities, and initiate each other into the reality of our own death such that we may live well. Heidegger suggests that this fourfold has been part of human dwelling from the beginning, characterizing our constructions such as homes and bridges as well as our agricultural projects and modes of social organization. On this analysis, true places are those that honor and remind us of the fourfold. As an example of such a place, Heidegger discusses an old bridge. The bridge respectfully allows the river to run its course while also allowing mortals to cross from one side to the other; it stands sturdily amidst changing weather; and its vaulting (and perhaps the statue of a saint perched upon it) evokes a sense of gratitude. Together, all this gathers the landscape into an intelligible place, one where we humans can feel at home. If any of these elements were absent, Heidegger suggests, then true dwelling would not be possible there; human beings would not feel at home.

Heidegger recognized that true dwelling was under threat already in 1951, when "Building, Dwelling, Thinking" was first published in German. He alluded to a housing crisis in Germany that prompted the construction of new residences, ones that were no longer rooted in the fourfold. In the flurry to construct many houses as quickly as possible, it seems that the builders opted not to create site-specific homes that would resonate with the natural landscape, but rather generic buildings that could placed here just as well as there or anywhere. With this mindset, Heidegger says, a new bridge would no longer be a site for dwelling, but rather "a mere something at some position, which can be occupied at any time by something else or replaced by a mere marker" (Heidegger, 1971, p. 153).

What is at stake when dwelling is under threat, when we build houses and bridges that are mere somethings at some position? Recall that dwelling with the fourfold is about the world having meaning, about feeling a sense of belonging, and about being inspired to live. It would seem better for a place to have these characteristics than to not. The language of the fourfold may seem abstract or antiquated, but I believe the fourfold is what people are pointing to when they speak of "feeling grounded" in a place or say that a place "has character" or "has history." Not all places are like this, as those of us who have traveled a lot will have realized. Compare the sprawling palimpsest of Rome to the dentist's-office flatness of Charlotte. Places without character or history—places that do not integrate the fourfold—may not encourage us to care for the earth or inspire us to strive for something deeply meaningful. They may also engender the sort of malaise that Charles Taylor (2007) writes about regarding the disappearance of the divine in the modern age, which I will discuss further below.

"Building, Dwelling, Thinking" is concerned with how the way we humans make structures is connected with our ways of being and thinking. Ignoring the fourfold, then, threatens us in many ways beyond the effects of our buildings and cities. To consider the elements of Earth and Mortals, for example, we might reflect on our civilization's reluctance to respond to climate change—to care for the earth that cares for us—and the way we tend to view death as a failure of life rather than a part of it. These attitudes are expressed in the recent behavior of certain tech-sector elites regarding space travel and anti-aging. For several years now, businessmen Jeff Bezos, Richard Branson and Elon Musk have been engaged in a "billionaire space race" (Lafranco, 2015) to install telecommunications satellites, establish extraterrestrial tourism, and eventually, the three of them hope, establish human colonies on Mars and/or the moon (Sky News, 2021; see also Tutton, 2021). These efforts rose to great fanfare in summer 2021, when Bezos and Branson each scrambled to launch themselves into space before the other (Maidenberg & Cole, 2021). Meanwhile, Bezos, Musk, Zuckerberg and others have been using their fortunes to fund research in anti-aging with the stated aim of "disrupting" death (Associated Press, 2016; TNN, 2017). A recent development to this end is the establishment of the Kempner Institute for the Study of Natural and Artificial Intelligence at Harvard University, which was funded through a donation by Zuckerberg and will be aimed in part toward better understanding and curing human disease (Powell, 2021). Such efforts paint a vision for a radically different mode of human being, one that defies the fourfold that has characterized humanity up to this point. As we will see in this paper, the metaverse is another effort along these lines.

With such dynamics unfolding, we might wonder about the prospects for humankind's continued dwelling. On this question, Heidegger suggests that a crucial step will be for us to see building as part of dwelling and both of these as "worthy of questioning and… worthy of thought" (p. 158). As I alluded to above, in fact, Heidegger contends that thinking itself belongs to dwelling along with building. This insight sheds new light on the threats to dwelling in our day: that thinking and dwelling are proceeding independently. Heidegger cautions that we must keep building and thinking together, and together with the fourfold:

> Building and thinking are… insufficient for dwelling so long as each busies itself with its own affairs in separation instead of listening to one another. They are able to listen if both—building and thinking—belong to dwelling, if they remain within their limits and realize that the one as much as the other comes from the workshop of long experience and incessant practice. (Heidegger, 1971, p. 158)

Put somewhat differently, Heidegger's concern here seems to be with scientific and technological advances that are inattentive to the qualities that make life worth living—focusing on the "can" of technological innovation instead of the "should." The upshot of this is that we human beings, as thinking creatures, "must ever learn to dwell" (Heidegger, 1971, p. 159). This involves both building and thinking out of a sense of dwelling, that is, with respect to the fourfold.

Heidegger's prognosis here reminds us that human dwelling is social and technological as much as anything. We are not just material creatures, and we do not live as monads. The technologies we create shape the ways we can dwell just as we shape them, to adapt a McLuhanesque phrase.[1] A concept that captures this social shaping, and which can be used to analyze ongoing currents in scientific and technological development, is that of social imaginary.

---

[1] The original quote is: "We become what we behold. We shape our tools and then our tools shape us." This quote is often attributed to Marshall McLuhan, but it was actually written by John Culkin. Still, it is certainly in line with Marshal McLuhan's thinking. See
https://mcluhangalaxy.wordpress.com/2013/04/01/we-shape-our-tools-and-thereafter-our-tools-shape-us/

# Social Imaginaries

We humans are social creatures; we must cooperate and foster relationships with others in order to survive and flourish. In *Being and Time*, Heidegger (1927/2010, §26) described this reality as our *being-with*. We also live in the past and future as well as the present; we ruminate and regret, and we plan and imagine. Heidegger's use of the term *understanding* includes this threefold temporality (§65); and on his account, human being is particularly oriented toward the future (I.5).

The term *imagination* can be used to describe our future orientation. With imagination, we reach beyond what is immediately available to our senses, and we can create and act upon our aspirations for the future. Research on human perception suggests that our imagination shapes what we take to be reality along with what is objectively present (de Lange et al., 2018). As we are social beings, it is unsurprising that some of our imaginings are social in nature; they have to do with our social reality. Such imaginings have been described with the concept of social imaginary.

A social imaginary is a set of symbols and values through which a group of people understand their social existence. Social imaginaries are "the ways people imagine their social existence [as] carried in images, stories, and legends" (Taylor, 2004, p. 23). As Charles Taylor (2004) writes, social imaginaries are at once descriptive and aspirational: They describe how a social group understands the world to be, and they direct that social group to further manifest its understanding. Social imaginaries also play a key role in the formation of social identity; they bind social groups together and help direct people's efforts. Taylor discusses a number of social imaginaries that have characterized modernity in the West (since ca. 1500): *the market*, or the idea that a set of equal, well-informed rational actors come together to exchange goods for mutual benefit; *the public sphere*, or the idea that political decision-making results from civil discussions wherein all stakeholders are equal participants; and so on. According to the theory, our belief in such social imaginaries in the first place is part of what makes them more and more the case.

The concept of social imaginaries has been applied to digital technologies to describe the ways such technologies and the attendant sociotechnical practices are developed and proliferate, as well as how desired changes in social life may be attained through advances in science and technology, with especial attention to the political and corporate influences involved (Jasanoff & Kim, 2015). For recent examples of research on sociotechnical imaginaries, see the recent special issue of *New Media and Society* on the theme (Mager & Katzenbach, 2021).

Potential social imaginaries are always emerging in the culture, as expressed through imagery, stories, mass media and the like. Only some of these will develop into a coherent social imaginary and penetrate widely throughout a society. The mechanism for this, according to Taylor (2004), is that certain social imaginaries are propounded by influential people. On Taylor's account, even earlier social imaginaries such as the market and public sphere began as ideas among a few elites and eventually spread out to broader strata. Nowadays, and focusing in particular on sociotechnical imaginaries, these influential elites tend to be business leaders in the technology sector, and they express their favored social imaginaries through advertising and public relations messages. The companies whose platforms are most widely and frequently used thus have the capacity to shape our social imaginations most deeply.

To recapitulate the discussion so far, social imaginaries are ways of thinking that configure how we dwell together as humans. While social imaginaries operate in all spheres of life, sociotechnical imaginaries are those particularly related to digital technology. These imaginaries shape our understanding of our current situation while also inspiring and guiding our actions into the future. In this paper, we will examine the metaverse as a particular sociotechnical imaginary, one that began to emerge in the 1990s and has perhaps reached a turning point in the past year.

## Enter the Metaverse

As mentioned in the introduction, the company Facebook changed its name to Meta in late 2021 as part of a pivot in its corporate strategy from social networking to the metaverse. The metaverse has been framed as a vision for the near future, a new understanding of our virtual environment and a natural extension of our social technology.

To be sure, Meta is not the only major company in the race to develop the metaverse, and nor was it the first. Months prior to Meta's announcements, Satya Nadella, CEO of Microsoft, promoted his company's vision to develop an "Enterprise Metaverse" primarily for business applications. And since at least 2018, Epic Games, purveyors of the game Fortnite and the software engine Unreal Engine, has been explicitly expanding its company's vision and offerings toward the development of the metaverse (Ball, 2019). (Even earlier, seeds of the metaverse concept can be seen in turn-of-the-century online role-playing games such as *Everquest* and *Second Life*.) Consonantly, these and other companies are gathering funding and leveraging investments in metaverse-relevant emerging technologies, such as virtual-reality headsets, augmented reality sensors, blockchain, and so on.

Commentators have been quick to point out the dystopian aspects of the metaverse vision. Expanding our digital world into a metaverse will not solve the issues of, say, misinformation and privacy, but likely exacerbate them (Sullivan, 2021). Moreover, the metaverse may be just a more effective enabler of our worst impulses of addiction and violence (Merchant, 2021). Such commentators point out that we need look no further than the term "metaverse" itself, which was coined by sci-fi author Neal Stephenson in his 1992 novel *Snow Crash*. In this book, the metaverse is a multimedia, computer-generated realm constituted by "The Street":

> a sprawling avenue where the buildings and signs represent "different pieces of software that have been engineered by major corporations." The corporations all pay an entity called the Global Multimedia Protocol Group for their slice of digital real estate. Users also pay for access; those who can only afford cheaper public terminals appear in the metaverse in grainy black-and-white. (Chayka, 2021)

As journalist Brian Merchant (2021) points out, the metaverse in *Snow Crash* was not an aspiration so much as a necessity in a world where material reality became unbearable.

These metaverse visions, if borne out, would seem to constitute a shift in human dwelling: what we understand to be our environment, how we relate to space, who we consider ourselves to be, what we take to be a life well lived… Of course, a shift of this kind has been underway at least since the invention of writing, which changed the meaning of distance and time. But we may be at a tipping point, given the fidelity and sociality of the digital experiences that the metaverse promises. In centuries' time, the birth of the metaverse could be seen as the completion of the information revolution, the moment at which the digital world took precedence over the physical one. (Indeed, that moment has already come for some of us.)

Even if the metaverse becomes our reality, the nature of that metaverse is not yet a foregone conclusion. We might envision, on one hand, the world from the 2008 Pixar film *WALL-E*, in which humans, pale and flabby, now live on a spaceship confined to glorified wheelchairs, each individually plugged into a digital environment, because the earth became uninhabitable. On the other hand, we might manage to develop a healthy "onlife" existence (see Floridi, 2015) where we engage with our digital lives in stewardship of our physical ones as well as the earth. This latter possibility, it should be clear, reflects true dwelling as described by Heidegger (1971), a way of being continuous with our human history.

At present, the metaverse is a social imaginary in the making. Commentary around the metaverse, though dating back to the 1990s, it still emerging and somewhat diffuse. Recall that social imaginaries are first formulated by a small group of elites—in the case of the metaverse, these elites are tech business leaders, venture capitalists, and certain journalists. Once the imaginary coheres sufficiently among these elites, it propagates throughout society as it continues to take shape in the public imagination. This is the stage of the metaverse at present. Recall in the weeks following Meta's announcement the myriad tech publication articles explaining what the metaverse is.

To better understand the precise nature of the metaverse as a social imaginary, the next section will analyze the vision for the metaverse being communicated by Meta. Building on prior research on the social imaginaries propounded by Facebook, this analysis shows how Facebook/Meta's corporate strategy has evolved since its founding.

Though there are still competing corporate visions for the metaverse as mentioned above (e.g., from Microsoft and Epic Games), I focus on Meta's vision in particular. This is because Meta's rebranding, newly dedicated funding for research and development, and recent public discourse all represent the largest investments in the metaverse to date. If we understand social imaginaries to be propagated by influential entities through channels such as advertising and public relations discourse, then Meta's vision is likely to be a driving force in the propagation of the metaverse social imaginary. What may we be in for, and what does this mean for human dwelling?

## Meta's Metaverse: A Thematic Analysis of an Emerging Social Imaginary

Facebook, founded in 2004, is currently the world's largest social network, with about 2.9 billion monthly active users as of mid-2021. Since 2007, Facebook has acquired or merged with dozens of other companies; some of these became integrated into the Facebook platform, while others, such as Instagram, WhatsApp and Oculus, have remained as separate brands (Wikipedia, 2021). Recognizing that the name "Facebook" was no longer appropriate as an umbrella term for such distinct offerings, the company changed its name to Meta Platforms in an announcement on October 28, 2021. As CEO Mark Zuckerberg explained as early as July 2021, he wants his company to be known not as "a social media company" but as "a metaverse company" (Newton, 2021).

After spending its first several years building up the Facebook platform, the company shifted to promulgating a vision for what it considers to be a better world. According to Haupt (2021), it did so through two key social imaginaries: first, that of *global connectivity* (ca. 2012–2015), and later, that of *global community* (ca. 2015–2017). The latter is expressed explicitly in Meta's mission statement, last modified in 2017: "Give people the power to build community and bring the world closer together." This understanding helps explain Meta's activities beyond its simply maintaining the Facebook network—for instance, its 2014 acquisition of Oculus, a virtual reality technology company. As Egliston and Carter (2020) demonstrate, Meta positioned Oculus as social media through its discourse, invoking themes such as: facilitating intimacy, affect and connectivity among users; connecting users to content creators; and seamlessly incorporating Oculus into everyday life. Such ambitions relating to "global community" beyond the Facebook platform have crystalized further in the company's name change to Meta and shift toward the metaverse, a new social imaginary.

To better understand how Facebook conceptualizes and communicates the metaverse as a social imaginary, I conducted a thematic analysis of of relevant public statements by Mark Zuckerberg in 2021. Methodologically, this study echoes the work of Haupt (2021), who studied Zuckerberg's public statements through 2017 in effort to discern the social imaginaries at play. This work is also resonant with that of Egliston and Carter (2020). As such, my work here can be seen as a continuation of the story told in these two contributions.

In my analysis, I included public statements by Zuckerberg regarding the metaverse from July 22 to December 16, 2021. I retrieved records of these statements from the Zuckerberg Transcripts collection within *The Zuckerberg Files* (Zimmer, n.d.). In all, 25 items were included in my analysis, ranging from short posts on Zuckerberg's Facebook page to long-form interviews and keynote transcripts. My analysis followed the standard guidance for thematic analysis in the social sciences as described by Braun and Clarke (2006). This involved reading the transcripts and/or viewing the videos several times through iterative rounds of inductive analysis. As Braun and Clarke write, the goal in thematic analysis is to develop themes that are internally homogenous and externally heterogeneous. Over my multiple rounds of analysis, I took notes on salient topics, continually compared the emerging themes for homogeneity/heterogeneity, and gradually arrived at a final list of themes.

In the end, six themes were identified: The metaverse is positioned as an **inevitable**, **industry-wide** development characterized by **presence** and **closeness** and designed in a **person-centered** way that will enable a new **economy**. I will discuss each of these in turn.

First, Zuckerberg describes the metaverse as a natural and **inevitable** evolution of the internet. As he explains, the internet began with text and soon incorporated still images, and now video is beginning to predominate. To some extent, this progression is a function of increased processing power and bandwidth, as well as better displays. Extrapolating this progress, Zuckerberg suggests that the future internet will be characterized by virtual reality (VR) and augmented reality (AR). Relatedly, just as no single company builds "the internet" but rather there are myriad "internet companies," the metaverse will not be built just by Meta; rather, Meta is just one corporate player in an **industry-wide** effort, if a catalyzing one. Zuckerberg states that his goal is to influence the market of VR/AR products in the same way that the first iPhone influenced smartphone design going forward. Specifically, he envisions a future in which VR/AR products are "more social," as he put it in a November 11 interview with Gary Vaynerchuk.

Two key themes that Zuckerberg uses to characterize the experience of the metaverse are presence and closeness. Zuckerberg uses the terms "embodied," "presence" and "shared" to describe the social space of the metaverse. Whereas today we look at the internet on our screens, in the metaverse we will be inside the internet, as Zuckerberg puts it, meaning we will feel more fully **present** with these new digital experiences and with each other in these social spaces. "The defining quality of the metaverse is presence, which is this feeling that you're really there with another person or in another place," Zuckerberg said in the 2021 second-quarter Facebook earnings call. Consonantly, interacting with the metaverse will not be done by typing and clicking, but rather through "natural" interactions such as gesture and eventually thought, and high-density displays will afford vivid and lifelike views. Next, Zuckerberg frequently discusses how the metaverse will **close** physical distances through "teleportation," which will be the metaverse analogue to today's clicking of hyperlinks. In the metaverse, people who are physically far away will be able to experience co-presence and "do almost anything you can imagine… with friends and family: work, learn, play, shop, create—as well as entirely new categories that don't really fit how we think about computers or phones today," Zuckerberg said in the October 28 Facebook Connect keynote.

Next, in Zuckerberg's vision the metaverse should be designed with the **person** as the central unit. On today's smartphone operating systems, apps are the central unit; and as a result, a person must create a separate account for each app, and our purchases, achievements, etc., cannot be carried between apps. This is equivalent, Zuckerberg says, to buying a jersey at a sporting event and not being allowed to wear that jersey outside the stadium. Meta's vision of the metaverse will not have this limitation. "Creation, avatars and digital objects will be central to how we express ourselves, and this is going to lead to entirely new experiences and economic opportunities," Zuckerberg said in the July 28 second-quarter Facebook earnings call. As such, the metaverse will enable new forms of commercialization and an entire **economy** of digital goods. Each user will have a Home Space where they can store their digital goods, and they will be represented by a customizable avatar. Zuckerberg hopes that "by the end of the decade that we can help a billion people use the metaverse and support hundreds of billions of dollars of digital commerce" (October 25, third-quarter earnings call). Advertising, he says, will "probably be a meaningful part of the metaverse."

For Meta, the metaverse represents a long-term vision that will be developed gradually in the coming decades; Zuckerberg says that in the next five to seven years we can expect to see technological implementations worthy of the metaverse name. He expects that gaming will be a major entryway for people to experience the metaverse, and he cites fitness and the workplace as two other domains already engaging with VR/AR technologies. Developing the metaverse will require surmounting a number of technical challenges, such as engineering varifocal lenses in VR systems to allow users to both read text up close and see clearly into the distance, and devising shared technical standards for interoperability that will allow digital assets to be shared across platforms and experiences. Meta and other companies are already investing research in these and other areas.

# The Metaverse as an Expression of the Ostrich Policy

As we begin to examine the prospects for human dwelling in the metaverse, it is worth considering the historical context in which Meta's name change and orientation toward the metaverse took place. Most broadly, these events occurred amidst extreme weather events, such as extreme wildfires in the western United States throughout 2021, which have sharpened concern about climate change, as well as an ongoing global pandemic that began in early 2020. More granularly, Zuckerberg's announcements came at a time of considerable chaos and frenzy for his company's reputation.

To be sure, Facebook has never been a stranger to scandal. Some of the platform's earliest issues had to do with academic research conducted using Facebook user data, raising concerns around privacy and personal agency (Hallinan et al., 2020; Zimmer, 2010). Such concerns were renewed with the Cambridge Analytica scandal in 2018, revealing the role Facebook data and analysis played in, most notably, the 2016 U.S. presidential election. Simultaneously it came to light that Facebook was used by Russian actors to influence the election, and soon after, Facebook was implicated in genocide in Myanmar. Along the way, public concern around the societal and mental-health effects of social media and algorithmic systems, as well as anti-trust issues with big tech firms, have been on the rise, evidenced in the creation and popularity of the 2020 documentary *The Social Dilemma* (Orlikowski, 2020). In October 2021, just weeks before the announcement of Meta's name change, the *Wall Street Journal* published a series of articles titled the *Facebook Files*, reporting on internal documents covering an array of issues related to user safety and mental health and revealing the company's internal tension between its knowledge of its wrongdoing and its aspirations for continued growth.[2]

In the opening lines of the 2021 Facebook Connect keynote, Zuckerberg acknowledged but immediately waved away these concerns. He said:

> I know that some people will say that this isn't a time to focus on the future. I want to acknowledge that there are important issues to work on in the present. There always will be. So for many people, I'm just not sure there ever will be a good time to focus on the future, but I also know that there are a lot of you who feel the same way that I do. We live for what we're building. And while we make mistakes, we keep learning and building and moving forward. For all of you who share these values, I dedicate today to you. In my mind, you are the heroes in our society who push the world forward. As long as I'm running this company, I will do my best to celebrate this spirit and absolutely go for it.

---

[2] The interested reader can turn to https://dayssincelastfacebookscandal.com, which has chronicled these and other scandals.

The emerging social imaginary of the metaverse suggests the development of a new universe for us to inhabit beyond our current one. The metaverse promises to, on one hand, reconstruct much of what exists in our universe, and on the other, go beyond certain limitations of our universe. In the background of this social imaginary is a turning away from our current world in many respects, and Zuckerberg explicitly presents the metaverse as a project to focus on rather than dealing with these other issues, from climate change to the pandemic to the present-day negative effects of his company's products. Simultaneously, as mentioned above, tech moguls are doubling down on space travel and health science research aimed at defeating death.

Heidegger (1971) warned of this sort of turning away in "Building, Dwelling, Thinking," as discussed above. Looking to the roots of these dynamics, we can consider the work of Hannah Arendt (1958), who was writing around the same time. In *The Human Condition*, Arendt observed that contemporary efforts to "escape from imprisonment to the earth" (she was commenting specifically on the launch of Sputnik the previous year) were part of the same desire as those around bioengineering and life extension, all of which were met with a kind of joy. Arendt suggests that it was humankind's capacity for abstract thought that opened the door to turning away from our present circumstances and toward scientifically imagined futures. In some ways, then, our contemporary sociotechnical imaginaries trace back to the scientific revolution, when our capacities for abstract reasoning came to a turning point. Yet Arendt believed that the 20th-century desire to "escape" from the prison of the earth was a contemporary shift, wrapped up in the dynamics of secularization. She writes:

> Should the emancipation and secularization of the modern age, which began with a turning-away, not necessarily from God, but from a god who was the Father of men in heaven, end with an even more fateful repudiation of an Earth who was the Mother of all living creatures under the sky? (Arendt, 1958, p. 2)

Digital technologies, which promised deeper social connection and possibilities, have proven to also come along with deeper alienation, loneliness, overwhelm, fear and nihilism. Over the past few years, and particularly during the pandemic, we have witnessed significant political strife and epistemological crises—and not to mention the looming threat of climate change. It is no wonder that Zuckerberg would rather ignore all this and instead build a new world to inhabit.

The tendency observed by Arendt (1958) and exhibited by Zuckerberg conforms to what Freud (1899/2010, p. 596) called "the ostrich policy": diverting one's attention from unpleasant information and focusing on something else. It is also understandable in light of secularization, as suggested by Arendt. Without a sense of a benevolent God, and without a sense that our very existence and the world itself are gifts, we may not be hopeful that the situation could improve. Moreover, with our scientific achievements we may believe that complete understanding of the universe is within reach and that a world of our own creation would be an improvement over the natural one.[3]

The spirit exhibited in the metaverse social imaginary "seems to be possessed by a rebellion against human existence as it has been given, a free gift from nowhere (secularly speaking), which he wishes to exchange, as it were, for something he has made himself" (Arendt, 1958, pp. 2–3). Arendt predicts that, should we manage to do so, it would not free us from physical labor as is so often promised, but rather transform "the whole of society into a laboring society" (p. 4), enslaving us to the machines we created. Here we may recall Zuckerberg's vision of the economy of the metaverse, in which many people will be "creators," selling digital experiences to others, all ultimately beholden to Meta's corporate policies.

But again, this particular future is not yet a foregone conclusion. Arendt writes: "The question is only whether we wish to use our new scientific and technical knowledge in this direction, and this question cannot be decided by scientific means; it is a political question of the first order and therefore can hardly be left to the decision of professional scientists or professional politicians" (Arendt, 1958, p. 3). In the spirit of discussing this "political question," we will now turn directly to our prospects for dwelling in the metaverse.

## Dwelling in the Metaverse?

In *A Secular Age*, Charles Taylor (2007) chronicles the emergence of exclusive humanism, a set of perspectives contending that God does not exist and that there is only the materialism of contemporary science. This perspective is a shattering novelty in the broad sweep of human history, 99.75% of which has been characterized by a default belief in divinities and "something beyond" the immediately evident (all but the past 500 years of our species' 200,000-year existence). Thus Taylor asks: can we feel at home in a godless place? Or will we always be tugging at the edges of the material, struggling for a glimpse of the transcendent? As we have seen, the question of divinities also figures in Heidegger's analysis of dwelling. We can imagine Heidegger asking, then, in addition to Taylor's question: can we feel at home in an earthless place? or in a skyless place? or in a deathless place?

The hope of technologists such as Mark Zuckerberg is that we will come to feel at home in the metaverse, that we will be able to flourish there as humans. In his media discourse, Zuckerberg has given a strong sales pitch for the benefits of the metaverse: creating more opportunities for social life, for economic development and freedom, for creative expression, etc. Yet if we are most human when we are dwelling with the fourfold, as Heidegger suggests, then it is worth examining the prospects for dwelling in the metaverse. Is it possible for the fourfold to gather in a computer-programmed world? Are there earth, sky, divinity and mortality in the metaverse? Or is it an earthless, skyless, godless and deathless place? If so, is Heidegger perhaps wrong, and true dwelling is possible even without these elements? Whatever our answers, will it be possible for us to truly feel at home in the metaverse?

---

[3] On these points, see Taylor (2007).

Drawing on Meta's public vision, the metaverse presents numerous threats to proper dwelling. We can consider these with respect to each element of the fourfold:

1. First, the metaverse constitutes a turning away from the **earth** as that which provides. The earth becomes a storehouse of resources to be exploited to fuel our experiences elsewhere, whether that is in a VR headset or outer space. Separated from their material origins, digital things may be experienced as unlimited and sempiternal. The meaning of geographic distance further erodes as people can "teleport," which creates a risk of focusing on faraway matters and not attending to nearby ones. Heidegger (1971) was concerned about a bridge becoming "a mere something at some position," and this concern reaches new heights in computer-generated spaces, as "earth" is generated apparently *ex nihilo*.
2. The metaverse also involves a turning away from the **sky**. In the metaverse, we are promised to have ultimate control over our environment. No longer will there be seasons or weather to contend with. Moreover, living in a programmed space suggests that anything we can imagine can be programmed, so there's no longer a "beyond" that defies our imagination or understanding.
3. We may also turn away from our **mortality**. The metaverse will allow us to interact with the avatars of people who have died and to present ourselves as perpetually young, drawing our attention away from the fact of our own death. We may also be tempted by promises of future technology for "mind uploading" that further discourage us from prioritizing our life projects given the limited and unknown amount of time we have. More prosaically, spending time in the metaverse may create issues with our physical bodies even as we become more "embodied" within the internet; one example is the eye strain and myopia that seem to accompany greater screen time and which may be further exacerbated with the prolonged use of VR headsets (BBC, 2020).
4. Finally, the metaverse involves a turning away from **divinity** as the source of vocation and the unexpected. In the metaverse, everything we experience will have been manmade, boxing out any space for the more-than-human. Consonantly, instead of seeing virtues such as humility and courage as guiding excellences to strive for, we will more readily see commercial objects and new consumable digital experiences as objects of desire.[4]

---

[4] On this point, see Vallor (2016), particularly her final chapter, "Knowing What to Wish For" (pp. 230–249).

More generally, the metaverse poses an additional threat to dwelling qua social imaginary. The more pervasive it becomes as a social imaginary, the less we will be able to see beyond it. Consider how social imaginaries such as the market and the public sphere hold sway over us today. It is difficult to see how economics could be possible without the notion of the market, or how politics could unfold without the public sphere, even if neither of these exists as cleanly as the social imaginary suggests. As a social imaginary, we can expect the metaverse to begin as an aspiration and gradually become reality through the way it shapes our efforts to make it become reality. If the metaverse poses threats to dwelling, then, these threats will be all the more pernicious once the metaverse becomes invisible to us,

To be sure, these threats of the metaverse seem to be extensions of threats to dwelling in the current age of the web. We already conceive of the earth in terms of its economic value as a stockpile of resources, something Heidegger (1977) observed decades ago; we already spend most of our time in climate-controlled built environments replete with artificial lighting; we already tend to ignore our mortality, and our posture and physical wellness are being threatened by sitting more than our ancestors and hunching over smartphones and laptops; and we already, as documented by Taylor (2007), overwhelmingly experience the world as manmade and do not believe in divinities. In this light, the metaverse will not necessarily introduce new threats to dwelling, but rather it will make many of our existing threats all the more inescapable.

The above analysis takes Meta's vision for the metaverse at face value with the goal of understanding the kind of future that Meta says it is seeking to build. However, there may be more to the story, whether Meta has ulterior motives or the public vision entails unforeseen consequences. Based on the history of Facebook, we should at the very least expect the latter. After all, even while Facebook publicly cultivated the imaginaries of global connectivity and global community through 2017 (Haupt, 2021), we now know that there was a whole suite of harms roiling beneath the surface. Thus many specific themes from Zuckerberg's vision could be called into question. Will Meta pursue the cooperation and interoperability necessary for this to be a truly industry-wide effort? Will the person really be the central unit of the metaverse, or the advertiser? Relatedly, how might advertising and profit motives undermine the utopian vision Zuckerberg has put forth?

Let us consider, for example, the first question. Zuckerberg has stated that Meta envisions the metaverse not just as the product of one company but an industry-wide effort, with the attendant standardization and cooperation. Now, that may be Zuckerberg's genuine intention; we cannot see into his heart. All we can do is look at his and Facebook's past actions and infer what we might be justified to expect in the future. With respect to this issue, Facebook's past actions suggest anti-competitiveness rather than cooperation. It is now well-documented that Facebook acquired could-be competitors such as Instagram knowing that they would pose a threat to the company, and the company has been increasing its spending on lobbying efforts (Li, 2021). There also seems to be a seed of deceit in the name Meta itself. If the metaverse is truly an industry-wide vision, why should one company have the name? It would seem that Meta wants to become synonymous with the metaverse, just as the term "Google" is now synonymous with "web search." We can imagine an alternative reality in which Zuckerberg did not rebrand Facebook but rather created Meta as a nonprofit consortium in which other tech companies had a say, along the lines of the Unicode Consortium.

Lastly, the way Zuckerberg has addressed the present-day harms of Facebook and Meta's other entities is telling. Over the years, Facebook has done very little in response to these harms, even if they were unintended and unexpected. When the metaverse inevitably causes unforeseen harms, how may we expect Meta to address them? For a clue, consider the timing of Meta's rebranding announcement, which came amidst a major expose series by the *Wall Street Journal* and after months of whistleblower reports, e.g., the publication of *An Ugly Truth: Inside Facebook's Battle for Domination* in July (Frenkel, 2021). Granted, the date for the Facebook Connect event was set far in advance, and no doubt the rebranding was in process for several months at least. Thus the timing of the announcement amidst this turbulence is likely coincidental. Even if coincidental, though, it is still irresponsible.

## How to Dwell in the Metaverse

As an inchoate social imaginary, the issues outlined above are not yet a foregone conclusion. The sort of future in which we will find ourselves with the metaverse depends on how it is designed, what technical challenges are surmountable and, perhaps most importantly, the extent to which it finds use and acceptance among the public. As the metaverse has not yet been developed, there is still time to ensure that it comes to be in a way conducive to human dwelling.

With all this in mind, I do see two major opportunities for improved human dwelling that the metaverse may present. If the metaverse is developed not as a "replacement" for the world—that is, not a separate universe that we "go into"—but rather as an additional layer of meaning atop our world as it exists, then the metaverse would constitute an invitation for us to care in new and additional ways. The seeds of this approach can be seen in existing AR games such as Pokémon Go, in which the real world is layered with an additional world of fantasy creatures to collect, journeys to be had and social interactions to be forged (see Holland & Denyer-Simmons, 2017). Games such as this offer new modes of human being-in-the-world as we discover hidden features in our world. Approaches such as this do not turn players away from the world, but rather more fully invest them in it. In addition to entertainment, Pokémon Go facilitates exercise, friendship and even community building. In the metaverse, future systems along these lines could build upon these prosocial aims.

Next, the metaverse could also provoke us to think more carefully about dwelling. On this point, Vella (2019) discusses how video games can lead us to critically engage with what it means to feel at home. Experiences in the metaverse may similarly prompt us to compare them with our experiences outside the metaverse, thus asking us to grapple with the meaning of dwelling. Recall Heidegger's point that thinking itself is part of dwelling, and that a central element of being human is that we "must ever learn to dwell" (Heidegger, 1971, p. 159). We may find that there are limits to forms of digital life that we, as the kind of beings we are, will accept, no matter how much technology corporations would like us to accept them.

As we move into the future, it would be useful to develop frameworks for ethical design in the metaverse. Such frameworks would communicate and cohere public values, guide designers and other stakeholders, and provide grounds for accountability. For decades, frameworks for ethical technology design have been devised in academic research; perhaps the most well-known example of these is Value Sensitive Design (see Friedman & Hendry, 2019). Over the past few years, this sort of work has begun to percolate both into the tech sector and the popular discourse, notably through the efforts of the Center for Humane Technology, founded in 2018. This work has engaged with a number of ethical values, such as privacy and freedom, as well as issues with the surveillance capitalism business model. However, these frameworks have not been linked directly to the issue of human dwelling, and they have not yet been adapted to the metaverse.

To close this discussion, then, I want to highlight a nascent framework that may be useful in designing technologies for the kind of metaverse conducive to human dwelling. Michael Heim's (1998) paper "Virtual Reality and the Tea Ceremony" reflects on the meaning of the traditional Japanese tea ceremony and the lessons it may provide for designers of virtual places who wish to help users feel more grounded and at home. Based on an analysis of the philosophy of the tea ceremony and his philosophical work on cyberspace, Heim identifies four principles of the tea ceremony particularly relevant to digital design, which are resonant with and perhaps more actionable than Heidegger's discussion of the fourfold:

- **Wa (Harmony)**:  The experienced world must cohere, i.e.., things must be part of a whole that hangs together. A cohesive experience is not just "one thing after another," but rather there is a sense to what happens. This principle suggests implementing constraints in user action and transitions between contexts.
- **Kei (Respect)**: We should acknowledge the presence of others, as well as the sacredness of the objects we use and the material components of our world. This suggests cultivating a caring attitude toward not only digital avatars and objects but also the nondigital world on which they rely.
- **Sei (Purity)**: In a spirit of minimalism, nothing should be wasted. This involves the discipline of constraint, and it runs counter to the majority of online experiences, in which "advertisers litter the void" (Heim 1998, p. 17). This suggests that, while advertising may have a role in the metaverse, it should not be all-pervasive, as it currently is on the web.
- **Jaku (Serenity)**: There must be ways to navigate the noise of online messages and to manage privacy. Specifically, there must be spaces of total privacy. Heim notes that this principle, along with Purity, is particularly fragile in the digital world.

The discussion of these principles has been rather limited; according to Google Scholar, as of January 2022, Heim's (1998) paper has only been cited 33 times, and there has been some popular-media discussion of Heim's ideas, such as on a 2015 episode of the *Buddhist Geeks* podcast. In light of the relevance of Heim's work in the context of the metaverse, I would suggest that these principles warrant renewed attention and further development. Incidentally, this suggestion is resonant with some recent work applying Japanese philosophy to AI ethics (see McStay, 2021).

## Conclusion

In this paper I have focused on the phenomenological concept of dwelling as a framework for examining how the metaverse social imaginary may impact human existence on a deep level. To me, dwelling is at the heart of human existence, and by addressing threats to dwelling, we may address any number of issues in passing. Still, it would be worthwhile to analyze other ethical and epistemological concepts threatened by the metaverse. This paper has obliquely made reference to privacy, autonomy, freedom and dignity, for example, all of which have been undermined in the past by Meta's products in the name of a business model based on surveillance capitalism. How likely is it that in the future, Meta's metaverse-centered products will not follow the same thread? What can be done before it is too late to ensure they do not?

As we continue to consider our prospects for dwelling with the metaverse, it may be instructive to return to Taylor's (2007) history of modern secularism. As Taylor suggests, it will take centuries to answer the question of whether humans can feel at home in a godless place. Our forms of social life have changed very quickly, *sub specie aeternitatis*, and these changes haven't had time to play out. Today we are seeing many troubling issues that on Taylor's account are side-effects of exclusive humanism (Taylor, 2007, ch. 8), such as the crisis of meaning and the attendant rise in mental health issues such as anxiety and depression. These may be indicative of the kind of future we are in for if we continue down the path we're on. It will take attentive development, then, to ensure that these are instead only bumps along the road to a better future.

## Declarations

**Consent for publication**: Not applicable
**Availability of data and material**: Not applicable
**Competing interests**: The author has no relevant financial or non-financial interests to disclose.
**Funding**: No funding was received to assist with the preparation of this manuscript
**Authors' contributions**: TG is the sole author if this manuscript.
**Acknowledgements:** To be provided after blind review.

<parsed value="\nHaupt, J. (2021). Facebook futures: Mark Zuckerberg’s discursive construction of a better world. *New Media & Society, 23*(2), 237–257. https://doi.org/10.1177/1461444820929315\n\nHeidegger, M. (1971). Building, dwelling, thinking. In A. Hofstadter (Trans. & Ed.), *Poetry, language, thought* (pp. 143–159). Harper & Row.\n\nHeidegger, M. (1977). The question concerning technology. In D. F. Krell (Ed.), *Basic writings* (pp. 307–341). Harper & Row.\n\nHeidegger, M. (2010). *Being and time* (J. Staumbaugh, Trans., & D. J. Schmidt, Ed.). State University of New York Press. (Original work published 1927)\n\nHeim, M. (1998). Virtual reality and the tea ceremony. In J. Beckmann (Ed.), *The virtual dimension*. Princeton Architectural Press.\n\nHolland, T., & Denyer-Simmons, H. (2017). *Embodied dwelling: The ontology of objects in Pokémon GO*. Paper presented at Australian and New Zealand Communication Association Conference 2017. http://dx.doi.org/10.17613/rjfw-4z80\n\nJasanoff, S., & Kim, S. H. (2015). *Dreamscapes of modernity: Sociotechnical imaginaries and the fabrication of power*. University of Chicago Press.\n\nLafranco, R. (2015, 13 April). Allen and Branson best Musk as the billionaire space race takes off. *Bloomberg*. https://www.bloomberg.com/news/articles/2015-04-13/allen-and-branson-best-musk-as-the-billionaire-space-race-takes-off\n\nLi, J. (2021). An ethical evaluation of corporate lobbying practices: A case study on Facebook Inc's lobbying strategies (unpublished B.A. thesis). University of Michigan.\n\nMager, A., & Katzenbach, C. (2021). Future imaginaries in the making and governing of digital technology: Multiple, contested, commodified. *New Media & Society, 23*(2), 223–236.\n\nMaidenberg, M., & Cole, D. (2021, August 21). Richard Branson and Jeff Bezos traveled to space. Here's how their trips differed. *Wall Street Journal*. https://www.wsj.com/story/photos-richard-branson-and-jeff-bezos-are-going-to-space-heres-how-their-trips-will-differ-e04c32dd\n\nMatney, L., & Hatmaker, T. (2021). Zuckerberg is turning trillion-dollar Facebook into a 'metaverse 'company, he tells investors. *TechCrunch*. https://techcrunch.com/2021/07/28/zuckerberg-is-turning-trillion-dollar-facebook-into-a-metaverse-company-he-tells-investors/\n\nMcStay, A. (2021). Emotional AI, ethics, and Japanese spice: Contributing community, wholeness, sincerity, and heart. Philosophy & Technology, 34, 1781–1802. https://doi.org/10.1007/s13347-021-00487-y\n\nMerchant, B. (2021, July 30). The metaverse has always been a dystopian idea. *Vice*. https://www.vice.com/en/article/v7eqbb/the-metaverse-has-always-been-a-dystopia\n\nNewton, C. (2021, July 22). Mark in the metaverse. *The Verge*. https://www.theverge.com/22588022/mark-zuckerberg-facebook-ceo-metaverse-interview\n\n\n21" />